\documentclass[conference]{IEEEtran}
\IEEEoverridecommandlockouts
\usepackage{cite}
\usepackage{amsmath,amssymb,amsfonts}
\usepackage{algorithm}
\usepackage[noend]{algpseudocode}
\usepackage{graphicx}
\graphicspath{ {./images/} }
\usepackage{textcomp}
\usepackage{xcolor} 
\usepackage{blindtext}
\usepackage{enumitem}
\usepackage{pgfplots}
\pgfplotsset{width=7cm,compat=1.8}

\usepackage{tikz,pgf}
\usetikzlibrary{calc,positioning,mindmap,trees,decorations.pathreplacing}
\usepackage{sverb}
\usepackage{caption}
\usepackage{verbatim}

\def\BibTeX{{\rm B\kern-.05em{\sc i\kern-.025em b}\kern-.08em
    T\kern-.1667em\lower.7ex\hbox{E}\kern-.125emX}}
\begin{document}

\title{Metamorphic IOTA}

\author{ 
\IEEEauthorblockN{Gewu Bu}
\IEEEauthorblockA{\textit{Sorbonne University,} \\
\textit{CNRS, LIP6}\\
F-75005 Paris, France}

 \and
 
\IEEEauthorblockN{Wassim Hana}
\IEEEauthorblockA{\textit{Sorbonne University,} \\
\textit{CNRS, LIP6}\\
F-75005 Paris, France}

\and
\IEEEauthorblockN{Maria Potop-Butucaru}
\IEEEauthorblockA{\textit{Sorbonne University,} \\
\textit{CNRS, LIP6}\\
F-75005 Paris, France}
}

\maketitle

\begin{abstract} 
IOTA opened recently a new line of research in distributed ledgers area by targeting algorithms that ensure a high throughput for the transactions generated in IoT systems. Transactions are continuously appended to an acyclic structure called tangle and each new transaction selects as parents two existing transactions (called tips) that it approves.  G-IOTA, a very recent improvement of IOTA, targets to protect tips left behind offering hence a good confidence level. However, this  improvement  had a cost: the use 
of an additional tip selection mechanism which may be critical in IoT systems since it needs additional energy consumption.  In this paper we propose a new metamorphic algorithm for tip selection that offers the best guaranties of both IOTA and G-IOTA. 
Our contribution is two fold. First, we propose a parameterized algorithm, E-IOTA, for  tip selection  which targets to reduce the number of random walks executed in previous versions (IOTA and G-IOTA) while maintaining the same security guaranties as IOTA and the same confidence level and fairness with respect to tips selection as G-IOTA. 
Then we propose a formal analysis of the security guaranties offered by E-IOTA against various attacks mentioned in the original IOTA proposal (e.g. large weight attack, parasite chain attack and splitting attack). Interestingly, to the best of our knowledge this is the first formal analysis  of the security guaranties of IOTA and its derivatives.
\end{abstract}


\section{Introduction}
 Bitcoin blockchain technology created a new design philosophy for executing and storing transactions in a decentralized and secure fashion \cite{nakamoto2008bitcoin}.
A blockchain is a distributed ledger that mimics the functioning of a classical traditional ledger (i.e.
transparency and falsification-proof of documentation) in an untrusted environment where the computation
is distributed. The set of participants to the system are not known and it varies during the
execution. Moreover, each participant follows his own rules to maximize its welfare.
Blockchain systems maintain a continuously-growing list of ordered blocks that include one or more
transactions that have been verified by the members of the system, called miners. Blocks are linked
using cryptography and the order of blocks in the blockchain is the result of a form of agreement
among the system participants. Participants strongly agree only on a prefix of the blockchain, the suffix of the blockchain may be different from one participant to another.

Bitcoin technology and similar proposals (e.g Ethereum) came with several drawbacks that prevent them from being used as  standard for IoT industry. In the field of IoT the main attributes that are concerned are the speed, scalability, and energy costs; all of which Bitcoin suffers from as limitations. Hence the introduction of IOTA \cite{popov2016tangle} designed specifically for the IoT industry. IOTA is a DAG (Directed Acyclic Graph) based distributed ledger, also known as the tangle,  aimed to overcome   limitations of Bitcoin when used in IoT environment while preserving equivalent security levels. IOTA uses tip selection algorithms for new transactions to approve two previous transactions.  IOTA suffers from certain limitations in terms of security and fairness with respect to approved transactions. Therefore G-IOTA \cite{bu2019g} was proposed as a new tips selection mechanism that combines a confidence fairness aware tips selection algorithm and a mutual supervision mechanism.  

In this paper we introduce a new approach, E-IOTA, that aims at maximizing the fairness level in tip selection by approving left behind tips, and improving confidence within the main tangle. E-IOTA randomizes tip selection  to reduce computational costs, as well as reduces left behind tips, and increases the security level of the tangle by avoiding a deterministic (predictable) tips selection algorithm (TSA). This makes the TSA and the tangle unpredictable for attackers. The algorithm creates a metamorphic main-chain that is as resistant to splitting attacks as IOTA and G-IOTA while reducing the costs of tip selection and hence preserving the energy of the nodes maintaining the tangle.

The organization of this paper is as follows. Section \ref{background} introduces IOTA and G-IOTA and identifies their drawbacks. Section \ref{sec:transaction} proposes E-IOTA that is designed to overcome the drawbacks of both IOTA and G-IOTA tangles. Section \ref{security} focuses on the security analysis of E-IOTA. 
Section \ref{performances} discusses the process of evaluation and testing of E-IOTA and provides the performance analysis and the comparison between the IOTA, G-IOTA and E-IOTA tangles. Section  \ref{conclusions} concludes the paper and discusses future research directions.

\section{Background on IOTA and G-IOTA}
\label{background}
In this section we present the design details of the IOTA system. Furthermore we focus its drawbacks  and describe the improvement G-IOTA and its drawbacks respectively.
\begin{figure}[h!]
    \centering
    \includegraphics[scale=0.30]{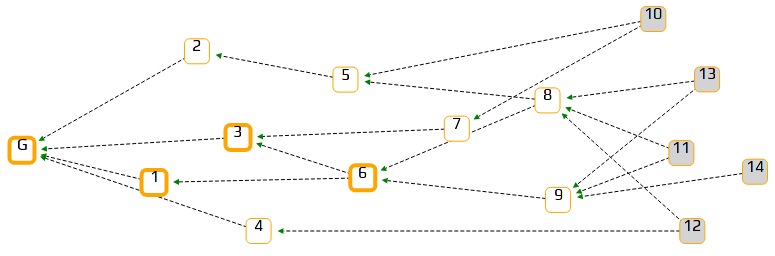}
    \caption{tangle example}
    \label{fig:my_label}
\end{figure}

\subsection{Tangle}
IOTA is a tangle-based distributed ledger. Unlike a standard blockchain it is a DAG where transactions are appended once verified. The genesis is the first and single transaction of the tangle. In the beginning of the tangle, there is an address with a balance that contained all of the tokens, in other words the genesis contains all the tokens that would be later on distributed to the network. The genesis transaction sent these tokens to several other founder addresses, addresses that are owned by the original investors of the IOTA project.



The transaction of the label G in Figure 1 is the genesis transaction of the tangle. The white boxes are approved transactions and the gray boxes are the unapproved transactions or tips. The tangle is characterized by its \emph{height} (the length of the longest oriented path to the genesis) and its \emph{depth} (the length of the longest reverse-oriented path to some tip).
Each square on the DAG in  Figure 1  represents a transaction on the tangle numbered from 1 to 14. When a transaction is generated in the network, two previous transactions must be approved. This approval is represented by  arrows in Figure 1. To illustrate, transaction 6 approves previous transaction(1) and transaction(3), so that it can be considered to be a valid transaction. Transactions can either directly or indirectly approve other transactions. A direct arrow between two transactions, for example transaction(6) and transaction(3), would be classed as a direct approval. If there is no direct arrow, but there is a path of lenght at least 2, as represented between transaction(9) and transaction(3) for example, then transaction(9) would be considered as an indirect approval of transaction(3). Newly issued transactions with no approvals are called tips. In order to best determine which two transactions to verify, IOTA uses what is called a tip selection algorithm (TSA) which will be further discussed in Section \ref{TSA}. 
\subsection{TSA}
\label{TSA}
The tip selection algorithm in IOTA tangle works on the basis that every incoming transaction has to approve two old transactions. In order to achieve that a walker is generated to traverse the tangle in a direction opposite of the path vectors based on a transition function until it reaches a tip. In the IOTA white paper  \cite{popov2016tangle} it is stated that an $N$ number of walkers are generated with every incoming transaction and race towards the tip, the first walker to reach a tip is the one to confirm it. In this paper for simplicity the N number of walkers is fixed to $N=1$, where there is no race between walkers, but rather one walker gets to traverse the path reaching a tip and confirming it. Taking into account that two old transactions need to be confirmed, therefore two walkers are generated asynchronously to reach these two tips using tip selection algorithms. IOTA's tip selections are categorized as follows:

\begin{description}[font=$\bullet$~\normalfont\scshape\color{black!50!black}]
\item [Uniform Random TSA] The uniform random tip selection algorithm is the simplest to implement since it chooses the two tips uniformly at random. The walker chooses paths  with equal probabilities until it reaches a tip.
\item [Logarithmic Markov Chain Monte Carlo TSA] With this algorithm, the number of children is more important than their weight \cite{bramas:hal-01716111}.i.e. the TSA prioritizes the chain with a higher number of children, rather than that with a higher weight.
\item [Weighted Markov Chain Monte Carlo TSA] This represents the main focus and being the most secure approach \cite{popov2017equilibria}. The details of this selections mechanism are provided below.
\end{description}

The weight of a transaction is directly correlated with the amount of work put in by a node in issuing it, for simplicity it is assumed that the weight of the transaction in this paper is fixed to 1. Transactions can also have what is known as a cumulative weight. Cumulative weight is the weight of a particular transaction plus the sum of the weights of other transactions that directly or indirectly approve it, for example in Figure 1 transaction(8) would be considered to have a cumulative weight of itself plus the weights of transaction(13), transaction(11), and transaction(12); hence the cumulative weight of transaction(8) would be equal to 4.

The Weighted TSA starts by initially putting a fixed number of walkers on the local DAG. Each walker performs a random walk towards the tips of the DAG with a probabilistic  transition function that depends on the cumulative weight of the site it is located to and its children. The more cumulative weight a next transaction has, the higher probability the random walker will go to it. Weighted Walk has a configurable parameter $\alpha$ to control the effectiveness of the cumulative weight \cite{popov2016tangle}, it is a parameter that determines the level of randomness the walker undergoes while choosing paths between chains. With a high  $\alpha$, even if the cumulative weights of two transactions have a small difference, the transaction with a bigger cumulative weight will have a higher probability to be chosen by the random walk. On the other hand, if $\alpha$ is small, even if the cumulative weights of two transactions have a big difference, they have almost the same probability to be chosen by the random walk.

\subsection{Confidence}
As a transaction receives additional approvals, its cumulative weight respectively grows making it more eligible to be chosen, hence within the tangle it is considered to have a high level of confidence. This helps in providing a more secure tangle by making it more difficult for the system to accept a splitting attack.
In a splitting attack, the attacker generates two conflicting transactions attached to previous transactions. 

It was stated in  \cite{popov2016tangle} that no rules are imposed for choosing which transactions a transaction will approve.
In order to issue a transaction, a node that wants to insert a new transactions does the following:
\begin{itemize}
\item the node chooses two other transactions to approve according to a tip selection algorithm. In general, these two transactions may coincide.
\item The node  checks if the two transactions are not conflicting, and does not approve conflicting transactions.
\item For a node to issue a valid transaction, the node must solve a cryptographic puzzle similar to those in the Bitcoin blockchain, in IOTA it is known as a proof-of-work approach where each node has to contribute to a small amount of work in the network by solving a light cryptographic puzzle. This allows them to issue transactions and add to the weight of the chain which increases security to the network.
\end{itemize}

A conflicting transaction is when a node manages to pass the same transaction twice in the chain, similar to a "buy one get one free" scenario where a client only pays for one item but gets two in return. The tangle then splits the chain to discard one of the conflicting transactions. Once the chain is split this gives the attacker the chance to flood both chains with other transactions keeping the weight of both chains balanced. This creates two validation paths where each transaction resides, and with this both transactions would be considered valid and non-conflicting.

The main rule that the transactions use for deciding between two conflicting transactions is the following: a node runs the tip selection algorithm many times, and sees which of the two transactions is more likely to be indirectly approved by the selected tip. For example, if a transaction was selected 97 times during 100 runs of the tip selection algorithm, we say that it is confirmed with 97\% confidence \cite{popov2016tangle}.

\subsection{Drawbacks}
\subsubsection{IOTA}
IOTA suffers from a trade-off problem of having a high $\alpha$  that leaves too many tips that do not receive confirmation in the tangle, or a low $\alpha$ that makes the tangle susceptible to attacks. As for left behind tips incoming transactions might disregard older transactions and approve newer ones, this taking into account lazy transactions that would approve closer transactions to avoid getting involved in heavy computations. This leaves many potential honest transactions forgotten and unapproved in the tangle that would later on get truncated. The best approach to reach all the tips in the tangle is by applying a uniform random walk since each point in the tangle has the same probability of being approved. This gets to question security, since a splitting attack would be easier to execute at that point, hence the need for the weighted MCMC walk. That approach would execute on certain paths with higher weights, and disregard other paths that might contain honest tips which would end up getting truncated. IOTA in order to achieve equilibria has to both be able to approve the majority of the tips while maintaining the security of the tangle \cite{popov2017equilibria}. With a high value of $\alpha$, we acquire more left behind tips, and with a low value of $\alpha$ we acquire a more vulnerable tangle to splitting attacks.
\newline \indent In terms of energy conservation and efficiency, a tangle of which its TSA performs weighted walks of a high value of $\alpha$ requires computation of the weights of each path in the chain; this would require higher energy in the tangle that would deem the tangle not too suitable for IoT transactions. This interferes with the concept of speed and energy efficiency on the scale of thousands of micro-transactions per second.
\newline \indent In terms of security the types of mitigation that IOTA implements to delude any types of malicious transactions entering the network require transactions being able to follow the path of the longest weighting chain. The more honest transactions the chain contains the higher the probability that they would approve other honest transactions. In order to achieve that the tip selection algorithm has to maintain a high $\alpha$ that would be able to distinguish between the weights of chains and pick the chain with the higher weight since it represents a higher confidence. An attacker in order to achieve a splitting attack would require splitting the chain at a certain time instant and maintaining the balance between the two chains with the probabilistic Uniform Random walk, since a transaction has the same probability on traversing on either side of the chain. In the studies of Bramas \cite{bramas:hal-01716111}, an IOTA tangle is susceptible to a splitting attack regardless of its TSA with the argument that later on a TSA would become deterministic, which refutes the reliability of the weighted tip selection algorithm.

\subsubsection{G-IOTA}
G-IOTA allows honest transactions in a tangle to increase their confidences evenly and quickly to meet the requirement of high confidence fairness \cite{bu2019g}. Implicitly, this way, no honest tips and transactions will be left behind during the growth of the tangle. Transactions with low confidence for a relatively long time can be considered as fake transactions.

G-IOTA is based on the weighted MCMC TSA. In addition, a Left-behind Tips Protection mechanism is integrated that allows tips that have been left behind regain the opportunity to be approved by incoming tips, which further decreases the transactions left behind. The new tips selection algorithm chooses not only two tips as the classical IOTA but may choose an additional tip, which is a left-behind tip in the tangle. That allows increasing the fairness in terms of transaction confidence for all honest transactions in tangle and guarantees the first approval for all honest tips.
\newline \indent Although the approach of G-IOTA tackles the issue of left behind tips that IOTA faces, the implementation of a third tip selection algorithm makes a transaction take more time and require more computation to reach approval in the tangle. This approach deviates from the purpose of an IoT based crypto-currency since it defies speed and energy efficiency more than that of the original IOTA.

\section{E-IOTA detailed description}
\label{sec:transaction}
This paper proposes E-IOTA a new approach for tackling left behind tips and maintaining high confidence levels in the tangle, see Algorithm \ref{alg1}.
E-IOTA has  as input three tip selection algorithms. Only one of these tip selection algorithms gets chosen randomly by a node with a certain probability each time it has to issue a transaction and a walker is generated. This would maintain balance in keeping a low $\alpha$ for tip confirmation, and high $\alpha$ in phases to contribute to the longest main chain which mitigates attacks.
 Tip selection algorithms input of E-IOTA are  as follows:
\begin{enumerate}
\item Weighted walks a relatively high $\alpha$, with probability $(1-p_2)$, making it very probable for the walk to go in the path of the weighted chain.\newline
Pros: limits the probability of an attacker being able to accomplish a successful splitting attack.
\item Weighted walks of a median $\alpha$, with probability $(p_2 - p_1)$ to lower the probability of going through a more highly weighted chain but rather give chance to confirm tips on other chains.\newline
Pros: Changes the main chain direction to deviate from a deterministic path and eliminate an attacker's predictability of the tangle.
\item Uniform random walks ($\alpha = 0$) with probability $p_1$ to roam to all sub-tangles and approve any left behind tips.\newline
Pros: Requires low computation where paths are probabilistically equal, this approach reduces computation demand in the tangle. It also maintains a high confirmation rate in the tangle by being able to explore all unapproved transactions.

\end{enumerate}
where $0 < p_1 < p_2 < 1$.
\begin{algorithm} [t]  \footnotesize
\caption{E-IOTA executed by a node $n_i$}
\label{alg1}
\begin{algorithmic} 
\State \%Two specific parameters for the algorithm are $p_1$ and $p_2$, such that $0 < p_1 < p_2 < 1$; $N$ is a  natural number. 
\State \%Each time when a node $n_i$ needs to chose a tip from the local tangle as the parent of its new transaction, it  draws a random number $r \in [0, 1)$.
\If{$r < p_1$}
\State $n_i$ generates $N$ uniform unweighted random walkers as TSA.
\ElsIf{$p_1 \leq r < p_2$}
\State $n_i$ generates $N$ weighted random walkers as TSA with a low $\alpha$ value.
\ElsIf{$p_2 \leq r < 1$}
\State $n_i$ generates $N$ weighted random walkers as TSA with a  high $\alpha$ value.
\EndIf
\end{algorithmic}
\end{algorithm}

E-IOTA reduces the need for the third walk proposed by G-IOTA, tackles the problem of left behind tips in IOTA, and reduces energy and computational power in the tangle making it more efficient. Section \ref{performances} describe the numerical results while running E-IOTA with specific values for $p_1$ and $p_2$.

\section{E-IOTA Security Analysis}
\label{security}
In this section we give a formal security analysis of E-IOTA. We look into all three mentioned attacks in the white paper of IOTA \cite{popov2016tangle}: \emph{Large Weight Attack}, \emph{Parasite Chain Attack} and \emph{Splitting Attack}. We prove that our fairness and effectiveness aware E-IOTA resists all of mentioned attacks.

\subsection{E-IOTA modelling}
According to the description of E-IOTA in section \ref{sec:transaction}, we define a model. We assume that the total average honest transaction rate is $\lambda$, that means that in a given time interval between $[t_0,t_0 + \tau]$, $\lambda$ transactions sent by honest users in the system will be in the tangle. For all $\lambda$ transactions that entered during $[t_0,t_0+\tau]$, $2 \lambda$ approvals are added into the tangle and approve old transactions already in the tangle. Among all these approvals, we can distinguish them into three classes:

1) $p_0$ percentage of them are chosen by Uniform Random Walks $TSA$, where $\alpha = 0$ i.e.\ all possible next hops have equal probabilities to be chosen by random walkers. To simplify, we assume that all the
tips will be chosen as parents by new coming tips with equal probability. 

2) $p_L$ percentage of them are chosen by Weighted Random Walks $TSA$ with a relatively low random $\alpha$, denoted by $\alpha_L$. With $\alpha_L$, random walkers will more likely to go to the next hop having higher cumulative weight, but others hops with less cumulative weight still have chance to be chosen.

3) $p_H$ percentage of them are chosen by Weighted Random Walks $TSA$ with a high $\alpha$, denoted by $\alpha_H$. With $\alpha_H$, random walkers go the next hops with the highest cumulative weight, with high probability.

Following the above classes, we consider that the $TSA$ in E-IOTA is a combination of  these three $TSA$s. The quantity of incoming approvals during a period of time, $A$, therefore consists of three types of approvals: $A_0$, approvals comming from tips chosen by unweighted random walks; $A_L$, computed from weighted walks with $\alpha_L$, and $A_H$ computed from weighted walks with $\alpha_H$. 
We assume that approvals sent come to the network during time interval $[t_0,t_0+\tau]$, will appear into the tangle in the time interval $[t_0 + \tau, t_0 + 2\tau]$ due to the network delay. Approvals come during the $[t_0,t_0 + \tau]$ therefore will not affect each other, because they are not yet   in the tangle. Each type of approval comes into the tangle therefore independently and only depends on the tangle in time $t_0$. According to the percentages of three $TSA$s, the $A$ can be defined as:





\begin{align} 
\label{fff1}
A = p_0 \times A_0 + p_L \times A_L + p_H \times A_H
\end{align}

By defining $A$ we proceed to analyse the three attacks.

\subsection{Large Weight Attack}

\begin{figure}
\centering
\includegraphics[width=0.45\textwidth]{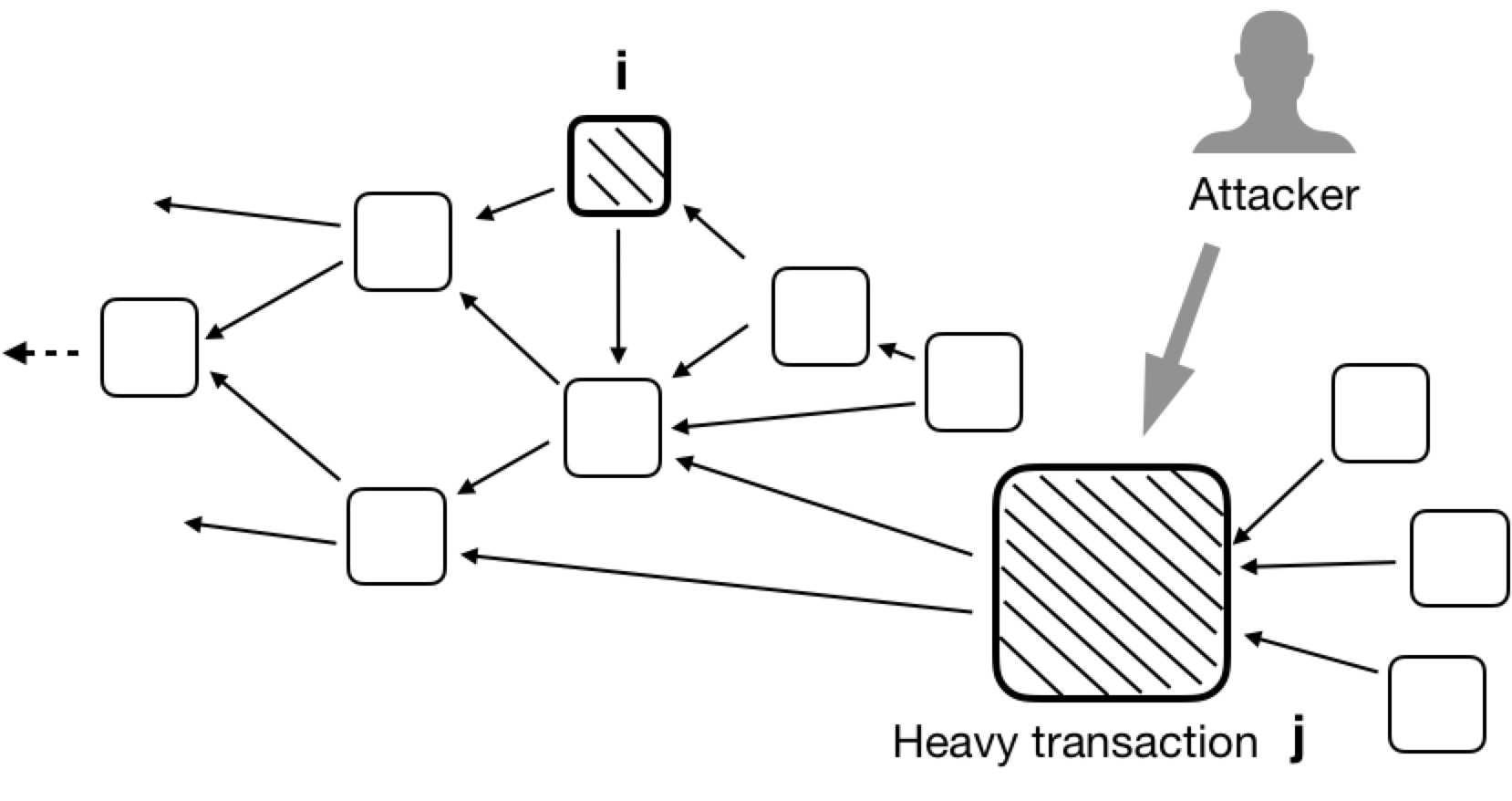}
\caption{Large Weight Attack}
\label{bat}
\end{figure}

\textbf{Description and Aim}
A large weight attack, shown in Figure \ref{bat}, occurs when an attacker tries to generate a "heavy" transaction, $j$, which conflicts with that of a previous transaction $i$, taking into account transaction $i$ has already been approved by several transactions and is considered as confirmed. That means money in $i$ is spent successfully. 
These two conflicting transactions do not belong to the same verification path. In this way an honest new coming transaction would not be able to approve both of them simultaneously. The attacker hopes that among new coming $A$, more approvals will approve $j$ than $i$ in the future, so the users will consider that $j$ is the correct transaction rather than $i$, so that the attacker can reuse the same money one more time, known as \emph{double spending}.

\textbf{Assumption}
In this case, we assume that $i$ has more approvals than $j$ at the beginning of the attack, as the transaction $j$ is just published onto the tangle and has not received any approval yet. Also we assume that $i$ has at least more than one tips approving it, because $i$ was already in the tangle for a while and received several approvals. Implicitly, $i$ has higher cumulative weigh than $j$.

We also take the assumption from the white paper of IOTA \cite{popov2016tangle} by limiting the maximal weight for any transaction. So that all transactions will have the same "heaviness". That means the attacker can only publish his transaction with the maximal allowing weight as same as others instead of sending a super "heavy" priority transaction.

\textbf{Proof}
For the $A_0$, they will more likely approve $i$ rather than $j$, as $i$ has more tips approving it than $j$ and each tips has equal probability to be chosen. Also as $i$ has higher cumulative weight than $j$, most of $A_L$ and $A_H$ will therefore approve $i$ rather than $j$. Therefore, in this case, all of the three types of $A$, will more likely to approve $i$ and to continue increasing the cumulative weight of $i$. That means, as time passes attacker's transaction $j$ has no chance to have more approvals than $i$.

Hence the large weight attack will not work in E-IOTA.

\subsection{Parasite Chain Attack}

\begin{figure}
\centering
\includegraphics[width=0.45\textwidth]{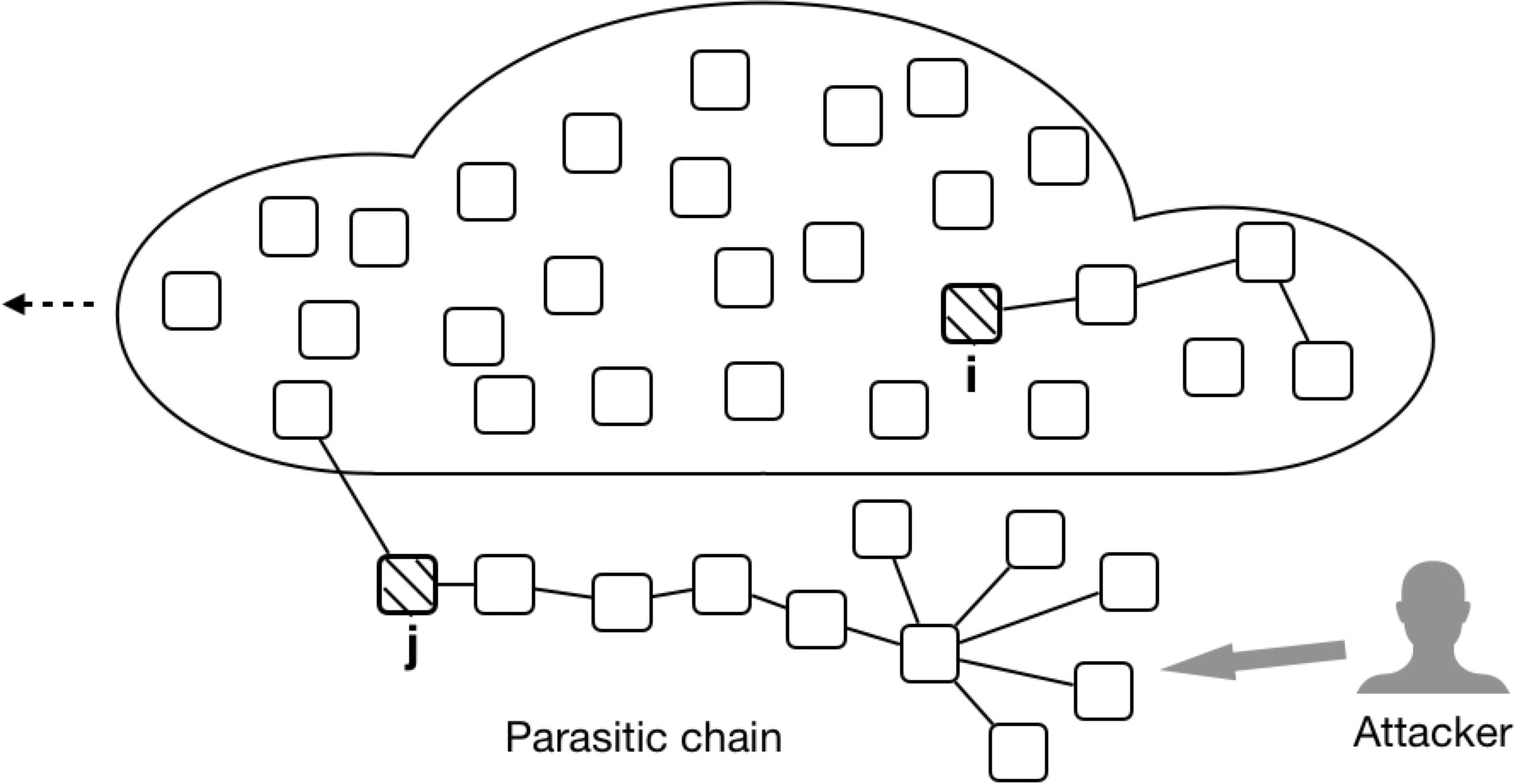}
\caption{Parasite Chain Attack}
\label{pat}
\end{figure}

\textbf{Description and Aim}
Parasite chain attack scenario is shown in Figure \ref{pat}. An attacker generates a sub-tangle (also called parasite chain) offline secretly. We indicate a transaction $j$ at the very beginning of this sub-tangle. $j$ approves transactions on the main tangle. After a time $T$, the attacker publishes a transaction $i$ into the main tangle conflicting with the transaction $j$ in the sub-tangle. The attacker continues working on the sub-tangle offline for a while making sure that the sub-tangle has as many as possible available tips to be chosen by random walkers. At the same time, $i$ may get several approvals in the main tangle and is confirmed (money spent). Then the attacker publishes his secret chain online and hopes that the more approvals in $A$ will approve $j$ than $i$ in the future to launch the double spending. 

\textbf{Assumption}
In this case, we assume that the computational power of the attacker is smaller than that of the honest users spent for publishing transactions. We also assume that the transactions approving $j$ in the sub-tangle are far more than that for $i$ as to show the worst case scenario present in \cite{bramas:hal-01716111}. Because although the computational power of honest users is higher than that of the attacker, it does not imply that all approvals from honest users will approve $i$, whereas on the other hand, all the computational power of the attacker will be used to generate approvals for $j$.

And also, we assume that all the transactions have the same weight.

\textbf{Proof}
As $j$ has more available tips, $A_0$ will more likely approve $j$ than $i$. In the worst case, we assume that $i$ does not receive any approval from $A_0$ with high probability. Hence most part of $A_0$ will approve $j$. However, as the computational power of the attacker is smaller than that of all honest users, the cumulative weight of sub-tangle published into the main tangle can not be higher than that of the main tangle after the time $T$. All the $A_H$ will therefore approve the main tangle, rather than the sub-tangle with a high probability. This implies that $A_H$ will not approve $j$. As long as the main tangle maintains a higher cumulative weight after the attacker attaches the sub-tangle, the new approvals from $A_H$ and $A_L$ will continue to contribute to the main tangle rather than the sub-tangle. Therefore, as long as $p_L + p_H > p_0$, which means the percentage of weighted walkers is higher than that of the unweighted random walkers, we can make sure that the majority of approvals contribute to the main tangle. $i$ will always receive more approve than $j$.

Hence the parasite chain attack will not work in E-IOTA as long as $p_L + p_H > p_0$.

\subsection{Splitting Attack}

\begin{figure}
\centering
\includegraphics[width=0.45\textwidth]{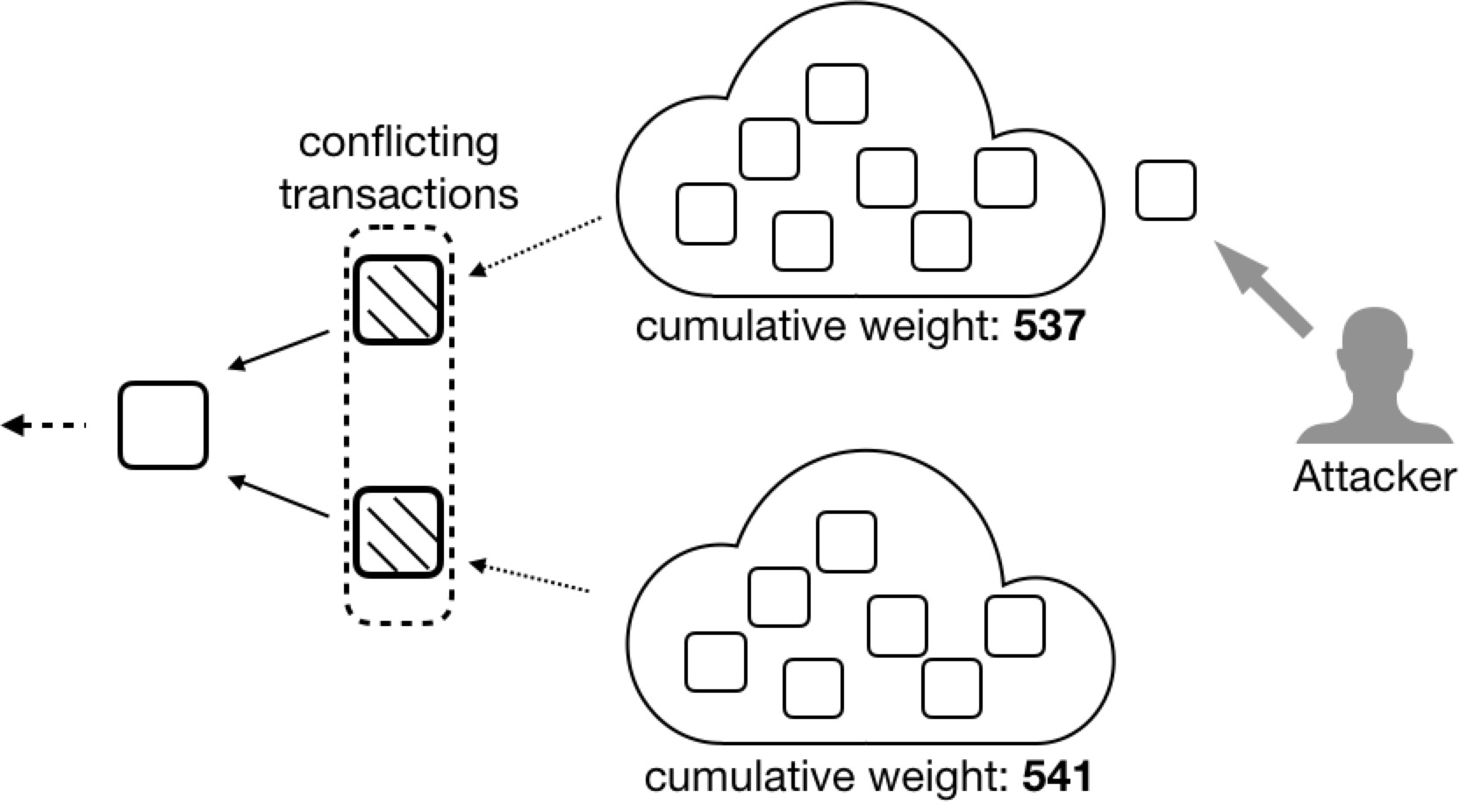}
\caption{Splitting Attack}
\label{sat}
\end{figure}

\textbf{Description and Aim}
In the splitting attack, shown in Figure \ref{sat}, an attacker will try to keep the balance of cumulative weights between conflicting branches. The conflicting branches can be generated by publishing two conflicting transactions, so that incoming transactions can only approve one of them. An attacker tries to maintain balance between the conflicting branches by publishing transactions to both branches while keeping them of almost similar weight. In this case the attacker will be able spend the same money in both conflicting branches. 

\textbf{Assumption}
We assume that the computational power of the attacker is $p_a\%$ of the total power of all the honest users $p$ spent for publishing transactions. And for this scenario, we assume that a splitting attack succeeds if the attacker has at least the same computational power as that of the computational power difference between honest users approving these two branches. Hence the attacker has to maintain the balance between the branches, in order to do so the attacker publishes his transactions to the branch that most of honest users won't approve, and hope that he can generate more approvals to keep the balance of cumulative weight between these two branches.

And also, we assume that all the transactions have the same weight.

\textbf{Proof}
The key problem represents itself in what is the maximal allowed $p_a$\% an attacker can have, so that E-IOTA can still mitigate the splitting attack. We know that among the $A$, $p_H$ of them will go to the branch with higher cumulative weight with high probability. We can therefore say that if $p_H > p_a$ then, no matter how hard the attacker tries, he can not generate more transactions for the branch with the lower cumulative weight, than the transactions generated by honest users towards the branch with the higher cumulative weight. This creates a significant difference between the weights of the branches with the weight of the branch that retains to the honest users remaining greater than that generated by the attacker. Hence, even approvals from $A_L$ are most likely to approve the heavier branch. The attacker therefore can not keep the balance between these two blanches.

The splitting attack therefore can not succeed as long as the $p_H > P_a$.

Note that to resist the lazy or cheating user mentioned in white paper \cite{popov2016tangle}, a \emph{Mutual Supervision} mechanism proposed in G-IOTA \cite{bu2019g}, can be added to E-IOTA to solve the problem.

\section{Performance evaluation}
\label{performances}
\subsection{Performance Evaluation Settings}
The main focus of this study is the comparison and analysis of IOTA, G-IOTA, and E-IOTA respectively. In the following we consider the same parameters as proposed for the evaluation of IOTA in \cite{tanglevisualization}:

\begin{description}[font=$\bullet$~\normalfont\scshape\color{red!50!black}]
\item IOTA running on a weighted MCMC TSA with $\alpha = 5$ 
\item G-IOTA running on a weighted MCMC TSA of $\alpha = 5$ 
\item VISA  transaction  network  can  process  2000tps \cite{koteska2017blockchain}; Therefore the three tangles were modeled to generate 2000tps reaching 8000 transactions.
\end{description}

For the E-IOTA algorithm described in Section \ref{sec:transaction}  the tips selections run with the following values for the parameters (these parameters show the best results for E-IOTA):
\begin{enumerate}
\item Weighted walks with $\alpha=5$  at probability $p=0.35$, making it very probable for the walk to go in the path of the weighted chain. We recall that this limits the probability of an attacker being able to accomplish a successful splitting attack.
\item Weighted walks of a median $\alpha$ of random value between 0.1 and 2 to lower the probability of going through a more highly weighted chain but rather give chance to confirm tips on other chains. This eliminates the eventuality for  an attacker to predict the tangle.
\item Uniform random walks of $\alpha=0$  with probability $p=0.1$ to roam to all sub-tangles and approve any left behind tips.  
\end{enumerate}

The three tangles (IOTA, G-IOTA and E-IOTA) are evaluated using the following metrics:

\begin{description}[font=$\bullet$~\normalfont\scshape\color{red!50!black}]
\item Number of approved transactions
\item Number of tips 
\item Confidence levels above 95\%
\item Number of Walks needed to reach the number of approved transactions
\item Simulation Speed
\end{description}

In our experiments clients run node.js v10.15.3 on Fedora 29 (x86-64) to generate the tangles IOTA, G-IOTA, E-IOTA respectively. We ran the experiments on three identical physical machines each using a different tangle, each machine equipped with:
\begin{description}[font=$\bullet$~\normalfont\scshape\color{red!50!black}]
\item Quad Core model: Intel Core i7-2600 processors running at 3.40 GHZ
\item Intel corporation 2nd generation Core processor Family integrated Graphics Controller
\item 8 GB RAM
\end{description}

The approach of having three identical machines is to be as unbiased as possible with the experimentation. This would assure that the three tangles are equally given the same computational power, and speed without giving the advantage to any. Calculating confidence between conflicting transactions would suggest running the random walk several times to determine the higher confidence between these transactions, and each transaction weight is given a fixed value of 1.  
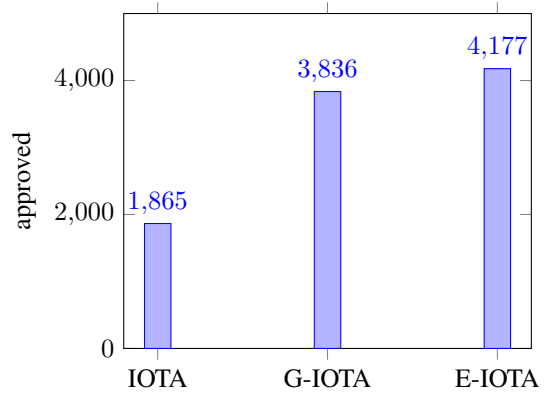
\begin{figure}[t!]
\centering    
\begin{tikzpicture}
\begin{axis}[
    ybar,
    ymin=0,
    ymax=5000,
    legend style={at={(0.5,-0.15)},
      anchor=north,legend columns=-1},
    ylabel={approved},
    symbolic x coords={IOTA,G-IOTA,E-IOTA},
    xtick=data,
    nodes near coords,
    nodes near coords align={vertical},
    ]
\addplot coordinates {(IOTA,1865) (G-IOTA,3836) (E-IOTA,4177)};
\end{axis}
\end{tikzpicture}
\caption{Number of approved transactions}
\label{fig:app}
\end{figure}

\subsection{Performance Evaluation Results}
The simulations for IOTA, G-IOTA, and E-IOTA were run 5 times respectively and the averaged results were taken.

\begin{figure}[t!]
\centering    
\begin{tikzpicture}
\begin{axis}[
    ybar,
    ymin=0,
    ymax=7000,
    legend style={at={(0.5,-0.15)},
      anchor=north,legend columns=-1},
    ylabel={tips},
    symbolic x coords={IOTA,G-IOTA,E-IOTA},
    xtick=data,
    nodes near coords,
    nodes near coords align={vertical},
    ]
\addplot coordinates {(IOTA,6135) (G-IOTA,4164) (E-IOTA,3823)};
\end{axis}
\end{tikzpicture}
\caption{Number of tips}
\label{fig:tip}
\end{figure}
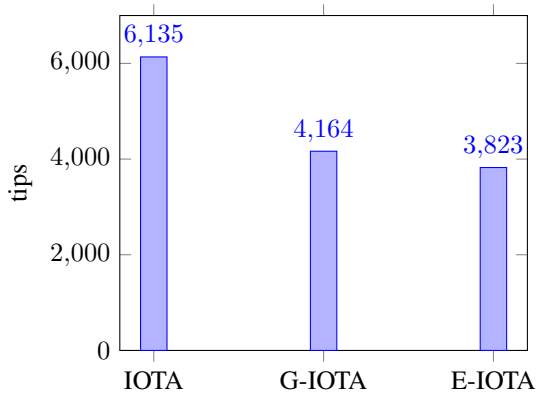

\begin{figure}[t!]
\centering    
\begin{tikzpicture}
\begin{axis}[
    ybar,
    ymax=10,
    ymin=0,
    legend style={at={(0.5,-0.15)},
      anchor=north,legend columns=-1},
    ylabel={confidence},
    symbolic x coords={IOTA,G-IOTA,E-IOTA},
    xtick=data,
    nodes near coords,
    nodes near coords align={vertical},
    ]
\addplot coordinates {(IOTA,7) (G-IOTA,8) (E-IOTA,7)};
\end{axis}
\end{tikzpicture}
\caption{ Number of transactions with Confidence higher than 95\%}
\label{fig:conf}
\end{figure}
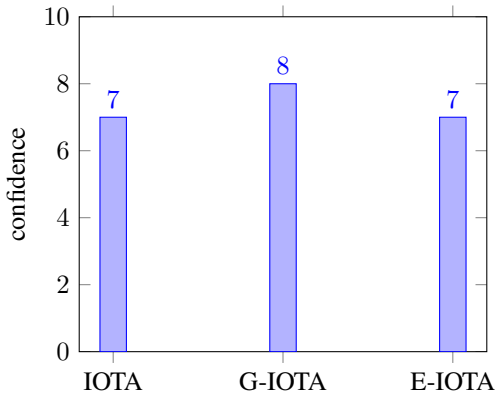

\begin{figure}[t!]
\centering    
\begin{tikzpicture}
\begin{axis}[
    ybar,
    ymin=0,
    ymax= 20000,
    legend style={at={(0.5,-0.15)},
      anchor=north,legend columns=-1},
    ylabel={walks},
    symbolic x coords={IOTA,G-IOTA,E-IOTA},
    xtick=data,
    nodes near coords,
    nodes near coords align={vertical},
    ]
\addplot coordinates {(IOTA,15998) (G-IOTA,15998) (E-IOTA,15998)};
\end{axis}
\end{tikzpicture}
\caption{Number of walks in the tangle}
\label{fig:walk}
\end{figure}
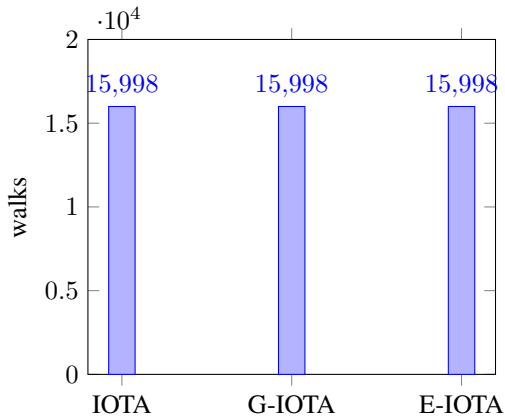

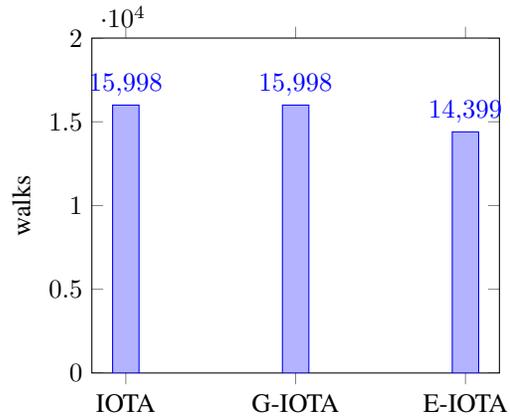
\begin{figure}[t!]
\centering    
\begin{tikzpicture}
\begin{axis}[
    ybar,
    ymin=0,
    ymax=20000,
    legend style={at={(0.5,-0.15)},
      anchor=north,legend columns=-1},
    ylabel={walks},
    symbolic x coords={IOTA,G-IOTA,E-IOTA},
    xtick=data,
    nodes near coords,
    nodes near coords align={vertical},
    ]
\addplot coordinates {(IOTA,15998) (G-IOTA,15998) (E-IOTA,14399)};
\end{axis}
\end{tikzpicture}
\caption{Number of walks that require computation}
\label{fig:walkcmp}
\end{figure}

\begin{figure}[t!]
\centering    
\begin{tikzpicture}
\begin{axis}[
    ybar,
    ymin=0,
    legend style={at={(0.5,-0.15)},
      anchor=north,legend columns=-1},
    ylabel={Time},
    symbolic x coords={IOTA,G-IOTA,E-IOTA},
    xtick=data,
    nodes near coords,
    nodes near coords align={vertical},
    ]
\addplot coordinates {(IOTA,92) (G-IOTA,130) (E-IOTA,87)};
\end{axis}
\end{tikzpicture}
\caption{Time elapsed for experiment}
\label{fig:time}
\end{figure}
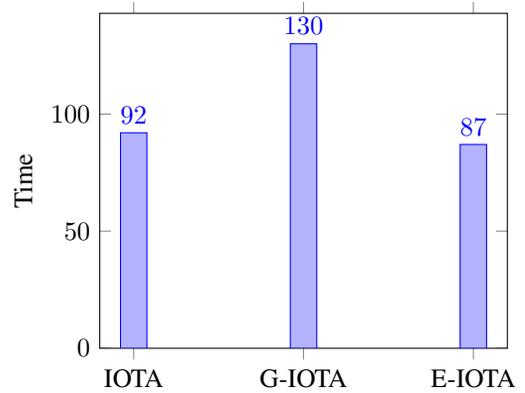 

In the following we  analyze the results of the experiment on the tangles of IOTA, G-IOTA, and E-IOTA. The high number of tips is to be taken into account that with a speed of 2,000tps reaching 8,000 transactions the last 2,000 transactions are most likely to be regarded as not approved yet in all three tangles. Therefore are considered as tips that managed to enter the network last. The analysis manages to differentiate between IOTA, G-IOTA, and E-IOTA respectively based on the number of approved transactions, the number of left behind tips, the needed walks to reach theses approved transactions, the needed computation, and the time elapsed for each tangle to reach these results. 

\subsubsection{Number of approved transactions}

In this section we compare the number of approved transactions for each tangle.
Figure \ref{fig:app} presents the number of approved transactions for the three tangles. The number of approved transactions for E-IOTA presents the highest value of 4,177 approved transactions whereas IOTA has only 1,865. The main reason as to why IOTA has a low approval value is due to the $\alpha$ value of 5. That would require walkers in the tangle to almost certainly go through the same paths disregarding other paths that contain transactions that have not been confirmed yet. In E-IOTA the option of exploring other paths exists which shows with the high number of approved transactions.
 
\subsubsection{Number of tips}
The number of tips as shown in Figure \ref{fig:tip} represents the lowest number of left behind retaining to E-IOTA 3,823 whereas the highest to IOTA of value 6,135. This is similar to the previous section as to due to the $\alpha$ value of 5 in IOTA a lot of transactions are not explored and end up becoming left behind. The probabilistic tip selection algorithm in E-IOTA with $\alpha=0$ or even the values between 0.1 and 2, gets to reduce this problem by traversing different paths and confirming the tips resulting in a low value of tips with E-IOTA.

\subsubsection{High confidence levels}
With respect to confidence levels  higher than 95\% the values in Figure \ref{fig:conf} show almost equal confidence levels for all three tangles. This represents a positive outlook on the aspect of security for maintaining high values of $\alpha$ to be able to mitigate splitting attacks. E-IOTA produces an $\alpha=5$   at a probability of 0.35, which helps it remain to preserve equal confidence levels in the tangle as compared to IOTA and G-IOTA.

\subsubsection{Number of Walks needed to reach the number of approved transactions}
The number of walks in the experiment represents the needed number of walks it took to reach the number of confirmed tips in IOTA, G-IOTA, and E-IOTA respectively.

Figure \ref{fig:walk} represents the number of walks of each of the tangles. The three tangles get to produce an equal number of walks of value 15,998. Comparing with figure \ref{fig:app}, IOTA managed to get 1,865 approvals ,G-IOTA managed to get 3,836 approvals, and E-IOTA managed to get 4,1177 approvals with these 15,998 walks. Figure \ref{fig:walkcmp} represents the number of walks that required computation of the weight of the tangle to reach these results of transaction approval. G-IOTA due to the fact that it manages to choose the third tip randomly, resulted in having equal values as that of IOTA at 15,998 instances the weight of the tangle was computed respectively with the number of walkers deployed. E-IOTA on the other hand only required 14,399 computations on the weight of the tangle due to the fact that some walkers are generated to follow the Uniform tip selection algorithm, which requires no need for the computation of the weight of the tangle. This represents a less need for walks to approve tips in the tangle and less computation for E-IOTA.

\subsubsection{Simulation Speed}
The speed of the experiment plays an important role in IoT micro transactions, the faster the ability to reach a certain point in time, (8,000 transactions in this experiment) is a necessary measurement in this experiment to remain on the context of an IoT based DAG.
Figure \ref{fig:time} shows that the least time needed to reach 8,000 transactions was that of E-IOTA with a time of 87 minutes, whereas IOTA came later with 92 minutes, and G-IOTA at 130 minutes. The fact is that IOTA requires a computation of the weight of the tangle each time a walker is deployed  consume time. G-IOTA similarly has to do the same but with the added third tip selection which lead to it being the slowest. E-IOTA manages to do so with occasional computation which explains as to why the simulation time for E-IOTA was smaller than that of IOTA, and G-IOTA.



\section{Conclusion and Future Research Directions}
 \label{conclusions}
In this paper we proposed E-IOTA which acts as a metamorphic  tip selection algorithm  that takes the advantages of each tip selection algorithm designed previously for IOTA or G-IOTA and runs them probabilistically in the tangle to overcome the drawbacks of both IOTA and G-IOTA. E-IOTA brings into attention the advantages of  metamorphic, self-adjusting tangles with respect to previous mono-strategy approaches. 
Our simulation results show that E-IOTA maintains a significant edge over IOTA and G-IOTA with respect to its performances, while maintaining equal level of confidence and security.
E-IOTA manages to tackle the problem of left behind tips that IOTA faces, it also manages to reduce computational energy, and approve more transactions. It manages to maintain the same security levels of that of IOTA and G-IOTA by using probabilistic tip selection mechanisms.  
 
Future research focuses on incorporating game theory, and establishing a self-aware tangle capable of punishing and rewarding nodes  based on performance and contribution to the tangle growth. The incorporation of game theory would bring more room to adding functionalities within the TSA. As long as the tangle is self-aware, it would adjust its structure based on its own detection of rising malicious activity in the tangle. Mechanism design techniques  can be used in order to punish  attackers that would  behave maliciously. A malicious transaction would have to play fair or get truncated. Another direction to be explored is to  run machine learning algorithms within the tangle in order to make it more self-aware.


\bibliographystyle{IEEEtran}
\bibliography{sample-dmtcs}

\end{document}